\documentclass[iop]{emulateapj}

\shorttitle{Lensing Magnification: A novel method to weigh high-$z$ clusters}
\shortauthors{Hildebrandt et al.}

\begin{document}

\title{Lensing Magnification: A novel method to weigh high-redshift clusters and its application to SpARCS}

\author{H. Hildebrandt$^{1,2,4}$, A. Muzzin$^3$, T. Erben$^4$, H. Hoekstra$^2$, K. Kuijken$^2$, J. Surace$^5$ L. van Waerbeke$^1$, G. Wilson$^6$, and H.K.C. Yee$^7$}

\begin{abstract}
We introduce a novel method to measure the masses of galaxy clusters
at high redshift selected from optical and IR Spitzer data via the
red-sequence technique. Lyman-break galaxies are used as a well
understood, high-redshift background sample allowing mass measurements
of lenses at unprecedented high redshifts using weak lensing
magnification. By stacking a significant number of clusters at
different redshifts with average masses of
$\sim1-3\times10^{14}M_\odot$, as estimated from their richness, we
can calibrate the normalisation of the mass-richness
relation. With the current data set (area: 6 deg$^2$) we detect a
magnification signal at the $>3$-$\sigma$ level. There is good
agreement between the masses estimated from the richness of the
clusters and the average masses estimated from magnification, albeit
with large uncertainties. We perform tests that suggest the absence of
strong systematic effects and support the robustness of the
measurement. This method - when applied to larger data sets in the
future - will yield an accurate calibration of the mass-observable
relations at $z\ga1$ which will represent an invaluable input for
cosmological studies using the galaxy cluster mass function and
astrophysical studies of cluster formation. Furthermore this method
will probably be the least expensive way to measure masses of large
numbers of $z>1$ clusters detected in future IR-imaging surveys.
\end{abstract}


\section{Introduction}
Galaxy clusters represent the largest collapsed structures in the
Universe which form in the peaks of the density field \citep[for
    reviews see]{RevModPhys.77.207,2011arXiv1103.4829A}. In order to
relate the observed cluster population to the density peaks seen in
numerical simulations the masses of the observed clusters have to be
inferred from observables. An accurate calibration of these
mass-observable relations is of great importance in astrophysical
studies of cluster-formation and -evolution as well as in cosmological
studies using the galaxy cluster mass function.

The richness of a galaxy cluster
\citep{1958ApJS....3..211A,1996AJ....111..615P,1999AJ....117.1985Y,2007ApJ...659.1106M,2009ApJ...703..601R}
is such an observable that can easily be measured from optical/IR data
out to high redshifts. The mass measurements that are required for
calibration, however, become increasingly difficult at higher
redshifts. The strong assumptions that are necessary to interpret
X-ray (hydrostatic equilibrium) or velocity dispersion (virial
equilibrium) measurements make it important to perform independent
cross-checks which do not rely on the same assumptions. This is
particularly important at

\vbox{ \vspace{0.5cm}\footnotesize \noindent 
$^1$~University of British Columbia, Vancouver, B.C. V6T 2C2, Canada\\
$^2$~Leiden Observatory, Leiden University, Niels Bohrweg 2, 2333 CA Leiden, The Netherlands \\
$^3$~Department of Astronomy, Yale University, New Haven, CT 06520-8101, USA\\
$^4$~Argelander-Institut f\"ur Astronomie, Auf dem H\"ugel 71, 53121 Bonn, Germany\\
$^5$~Spitzer Science Center, California Institute of Technology 220-06, Pasadena, CA 91125, USA\\
$^6$~Department of Physics and Astronomy, University of California-Riverside, 900 University Avenue, Riverside, CA 92521, USA\\
$^7$~Department of Astronomy and Astrophysics, University of Toronto, 50 St George Street, Toronto, Ontario, M5S 3H4, Canada
}

\noindent high-$z$ where clusters are dynamically younger and
equilibrium assumptions become increasingly questionable.

One way to measure mass that does not rely on assumptions like
hydrostatic or virial equilibrium is weak gravitational lensing
\citep[WL; for a comprehensive review see][]{2001PhR...340..291B}. The
commonly used measurement of the shearing of galaxies breaks down for
higher redshifts because the sources cannot be resolved anymore and
hence their ellipticities cannot be measured. This is particularly
true for ground-based data and lens redshifts approaching
$z\sim1$. Those higher redshift regions are, however, of special
interest for cluster astrophysics because they approach the cluster
formation epoch where observational guidance is crucial. The
measurements of dark energy using clusters \citep[see
    e.g.][]{RevModPhys.77.207} as well as measurements of primordial
non-Gaussianity \citep[e.g.][]{2011PhRvD..83b3015M} benefit from
a large redshift baseline including high-$z$ observations.

Here we present a first measurement of the average mass of such
high-$z$ cluster lenses based on the magnification effect of WL as
proposed by \cite{2010ApJ...723L..13V}. This method does not require
one to resolve the sources and hence can be applied at higher
redshifts. By stacking a number of clusters we can measure their
average mass. The clusters are selected from the Spitzer Adaptation of
the Red-sequence Cluster Survey
\citep[SpARCS\footnote{\url{http://faculty.ucr.edu/$\sim$gillianw/SpARCS}};][]{2009ApJ...698.1943W,2009ApJ...698.1934M,2010ApJ...711.1185D};
whereas the background sources are selected from catalogues of the
Canada France Hawaii Telescope Lensing Survey (CFHTLenS), which are
based on imaging data from the CFHT Legacy Survey (CFHTLS), via the
Lyman-break technique
\citep{1996ApJ...462L..17S,2002ARA&A..40..579G}. The data are
presented in Sect.~\ref{sec:data}. In Sect.~\ref{sec:method} we
describe the magnification method. Results are reported in
Sect.~\ref{sec:results} and discussed in
Sect.~\ref{sec:discussion}. Throughout we assume a flat $\Lambda$CDM
cosmology with $\Omega_{\rm m}=0.27$, $\Omega_\Lambda=0.73$, and
$H_0=70{\rm km}^{-1}{\rm s}^{-1}{\rm Mpc}^{-1}$ according to
\cite{2011ApJS..192...18K}. All magnitudes are in the AB system.

\section{Data set}
\label{sec:data}

The cluster sample used for this analysis is drawn from the 9 deg$^2$
XMM-LSS field which is one of six fields in the 42 deg$^2$ SpARCS
survey.  Clusters are selected as overdensities in a combined color
and position space using the cluster red-sequence method developed by
\cite{2000AJ....120.2148G,2005ApJS..157....1G}.  SpARCS uses a
z$\prime$ - 3.6$\micron$ color selection which allows the detection of
clusters up to redshifts as high as $z \sim$ 1.5.  Photometric
redshifts for the clusters are calculated based on the color of the
cluster red-sequence and have been refined using a subsample of 33
clusters with confirmed spectroscopic redshifts.  Comparison with the
spectroscopic sample shows that the photometric redshifts are accurate
to $\Delta z$ $\sim$ 0.1 up to $z \sim$ 1.4. The richness is
  estimated from 3.6$\mu$m data. Since the $K-$3.6$\mu$m color of
  galaxies is independent of star formation history and redshift for
  $z \la 2$, these 3.6$\mu$m richness estimates are equivalent to
  $K$-band ones. Further details of the SpARCS cluster catalogs can
be found in \cite{2009ApJ...698.1934M} and \cite{2009ApJ...698.1943W},
and a more detailed discussion of the cluster detection algorithm can
be found in \cite{2008ApJ...686..966M}.

In the following we concentrate on secure clusters (SpARCS flux
$>9$). This ensures that only massive systems and no potential false
positives are stacked. Selecting clusters in this way leads to more
massive systems at high redshift than at low redshift, but with the
limited data set used here we cannot afford more physical cuts
(e.g. fixed richness). Within the overlap area between SpARCS and
CFHTLS we find 48 moderately rich and securely detected clusters with
redshifts between $z=0.2$ and $z=1.6$.

As sources we use Lyman-break galaxies (LBGs) selected from catalogues
of the CFHTLS field W1 which overlaps with the SpARCS XMM-LSS field by
approximately six deg$^2$. The optical multi-color catalogues are
created from stacked images \citep[see][for details on the imaging
  data reduction with the THELI pipeline]{2009A&A...493.1197E} in a
similar way as described in \cite{2009A&A...498..725H} with the
notable exception that the PSF is brought to the same Gaussian shape
over the whole pointing and for all five images in the
$ugriz$-filters of each field. This allows more accurate color
measurements which results in better photo-$z$'s and a more accurate
Lyman-break galaxy selection. Details of this method will be presented
in a forthcoming paper. The LBG color selection of $u$-, $g$-, and
$r$-dropouts is identical to the one presented in
\cite{2009A&A...498..725H} with the exception that $u$-dropouts are
only required to have $u-g>1$ instead of $u-g>1.5$. This relaxed $u-g$
cut is identical to the one used by \cite{2010A&A...523A..74V} whose
luminosity function measurements we use for calibration.\footnote{This
  change was introduced to make the sample more comparable to other
  $u$-dropout studies, like \cite{1999ApJ...519....1S}.}

In the following we use $z\sim3$ $u$-dropouts and $z\sim4$
$g$-dropouts. The $u$-dropouts are the cleanest background sample in
terms of low-$z$ contamination and they are the lowest-redshift
dropout sample detectable from the ground which results in the highest
apparent brightness. The $g$-dropouts are fainter by $\sim0.5{\rm
  mag}$ due to their larger distance resulting in larger photometric
errors. They also show larger low-$z$ contamination. Thus, we can only
safely cross-correlate them to clusters with intermediate and high
redshifts - beyond the redshifts of the possible contaminants. In
  the following, we correct for the dilution due to contamination
  (boosting the signal by $10\%$) whenever $g$-dropouts are used. The
CFHTLS-Wide data are too shallow to select $r$-dropouts in significant
numbers.

\section{Mass measurement}
\label{sec:method}

The signal-to-noise ratio (S/N) of the lensing signal per background
galaxy is generally lower for magnification-based methods than for
shear-based ones.\footnote{This depends on the slope of the number
  counts \citep[see e.g.][]{2009A&A...507..683H}. It is beneficial to
  select a source population with a steep slope.} However, since
magnitudes - the only requirement for magnification - are easier to
measure than shapes, there are always more galaxies and in particular
higher-redshift galaxies available for magnification. Thus, there is a
break-even redshift for each data set beyond which magnification
becomes more powerful than shear because ellipticity measurements
become impossible. This is the case with clusters at $z\ga0.8$ where
very few galaxies with reliable shape measurements are available at
redshift higher than the cluster redshift, as it is very difficult to
measure shapes of galaxies with $z>1$ from ground-based data.

Since the S/N is too low for a single high-$z$ cluster to be detected
via magnification with the background samples described above we rely
on stacking the signals of several clusters. In this way we can  in principle estimate their average mass, a method very similar to
what is usually done in galaxy-galaxy-lensing. Therefore, the
estimator is identical to Eq.~11 in \cite{2009A&A...507..683H},
i.e. we look for correlations in the positions of LBGs and the SpARCS
clusters. Those positions are correlated due to the magnification-bias
effect of WL.

The magnification signal scales linearly with the slope of the number
counts, $\alpha$.\footnote{The $\alpha$ we use here should not be
  confused with the faint-end slope of the luminosity function which
  is also often called $\alpha$.} Thus, it is important to select a
source population with a large $\left|\alpha\right|$. For the
interpretation of the signal, $\alpha$ needs to be measured.  However,
we cannot use the measured $\alpha$ directly because of incompleteness
in our catalogue. Furthermore, the incompleteness changes over the
magnitude bin and so does the $\alpha$ itself. In order to account for
these effects we model the incompleteness as a function of magnitude
by comparing the LBG number counts in the CFHTLS-Wide to the ones in
the CFHTLS-Deep \citep{2009A&A...498..725H}. We then multiply the
luminosity function of the LBGs measured in
\cite{2010A&A...523A..74V}\footnote{Note that the volume probed by
  each of the dropout samples in the four deg$^2$ of the CFHTLS-Deep
  fields is $\sim0.2{\rm Gpc}^3h^{-1}$ so that cosmic variance on the
  number counts is negligible. For comparison: the Millennium
  Simulation has a volume of $0.125{\rm Gpc}^3h^{-1}$.  } with this
incompleteness function. Next we artificially magnify the same
luminosity functions for different values of the magnification $\mu$
and again multiply with the un-magnified incompleteness function. Then
we compare the numbers of LBGs in a chosen magnitude bin between the
un-lensed and the lensed case. A linear fit to the number excess (or
depletion) as a function of $\mu$ yields $\alpha_{\rm eff}$, the
effective $\alpha$ to be used in the interpretation of the
measurement.

The two problems of varying incompleteness and varying $\alpha$ over
the magnitude bin should be considered in all magnification
applications. Just correcting the number counts for incompleteness
before measuring $\alpha$ and then averaging the individual $\alpha$'s
of all galaxies (assigned depending on their magnitudes) in the
magnitude bin only works if the changes of the incompleteness as well
as of $\alpha$ over the magnitude range are moderate. The $\alpha_{\rm
  eff}$'s that we use here can easily differ from these naive $\alpha$
estimates by $20$-$30\%$.

Besides estimating accurate $\alpha_{\rm eff}$'s it is important to
cleanly separate the lenses and the sources in redshift to prevent
physical cross-correlations from biasing the WL measurement. Using
LBGs as a background sample (and $u$-dropouts in particular) has the
great advantage that they represent a very clean high-$z$ sample and
their small amount of low-redshift contaminants are very well
understood \citep{1999ApJ...519....1S,2003ApJ...592..728S}. In this
way we can make sure that no galaxies with redshifts comparable to the
cluster redshifts are included in the background samples and that
physical cross-correlation (in contrast to magnification-induced
cross-correlations) do not play a role here.

The cross-correlation signal $w(\theta)$ is then related to the
magnification $\mu$ through \citep{2001PhR...340..291B}:
\begin{equation}
w(\theta)=\left<\alpha_{\rm eff}-1\right>\,\delta\mu(\theta)\,,
\end{equation}
with $\delta\mu(\theta)=\mu(\theta)-1$. The $\left<\alpha_{\rm
  eff}-1\right>$ for the bright LBGs selected from the CFHTLS-Wide,
which we use here, is always positive, so that we expect a positive
cross-correlation signal.

Since the clusters span a range of redshifts and masses - as suggested
by their richness - we cannot just fit a simple cluster mass model
like a singular-isothermal-sphere (SIS) to the stacked signal and
  estimate the average cluster mass directly. Rather we fit the WL
signal with a one-parameter model which is constructed from a weighted
sum of SIS models for all $N$ clusters:
\begin{equation}
\label{eq:multi_SIS}
\mu_{\rm comb}(a)=\sum^N_{i=0}\mu_{\rm SIS}(z_i;a\,M_{200,{\rm richness},i})\,.
\end{equation}

The relative weighting between the clusters in this model takes the
different redshifts $z_i$ (and hence different lensing efficiencies
and different critical densities) as well as the different masses
$M_{200,{\rm richness},i}$, as estimated from their richness via
  Eq.~9 of \cite{2007ApJ...663..150M}, into account. Hence the fit
  does not directly yield an average mass. Instead the normalization
  of the mass-richness relation is fitted under the assumptions that
  the slope of this relation does not change with redshift, the
  redshifts for the clusters are accurate, and that the background
  cosmology is known.

The magnification for an SIS takes the form:
\begin{equation}
\mu(\theta)=\frac{\theta}{\theta-\theta_{\rm E}}\,,
\end{equation}
with $\theta_{\rm E}=4\pi\left(\frac{\sigma_v}{c}\right)^2\frac{D_{\rm
    ds}}{D_{\rm s}}$ being the Einstein radius of an SIS with velocity
dispersion $\sigma_v$; $D_{\rm s}$ and $D_{\rm ds}$ are the angular
diameter distances from observer to source and from deflector to
source, respectively. The velocity dispersion is related to the mass,
$M_{200}$, by
\begin{equation}
\sigma_v^3=\frac{1}{2^{3/2}}M_{200}\sqrt{\frac{4}{3}\pi200\rho_{\rm crit}(z)G^3} \,,
\end{equation}
where $\rho_{\rm crit}(z)$ is the critical density at the cluster
redshift, $z$, and $G$ is the gravitational constant. Here we estimate
the masses with the mass-richness relation at $z\sim0.3$ from
\cite{2007ApJ...663..150M} which is based on dynamical mass estimates.
A fitted $a$ of unity would mean that the mass-richness relation at
the redshifts of our cluster is identical to the one from
\cite{2007ApJ...663..150M}.  The mass estimated from magnification
becomes $M_{\rm magnification}=a\,M_{\rm richness}$\,.

In the future, larger cluster samples will allow us to bin the
clusters much more finely so that an equal weighting can be used and
no external mass-observable relation is needed for proper scaling.

\section{Results}
\label{sec:results}

\begin{figure*}
\includegraphics[width=0.32\textwidth]{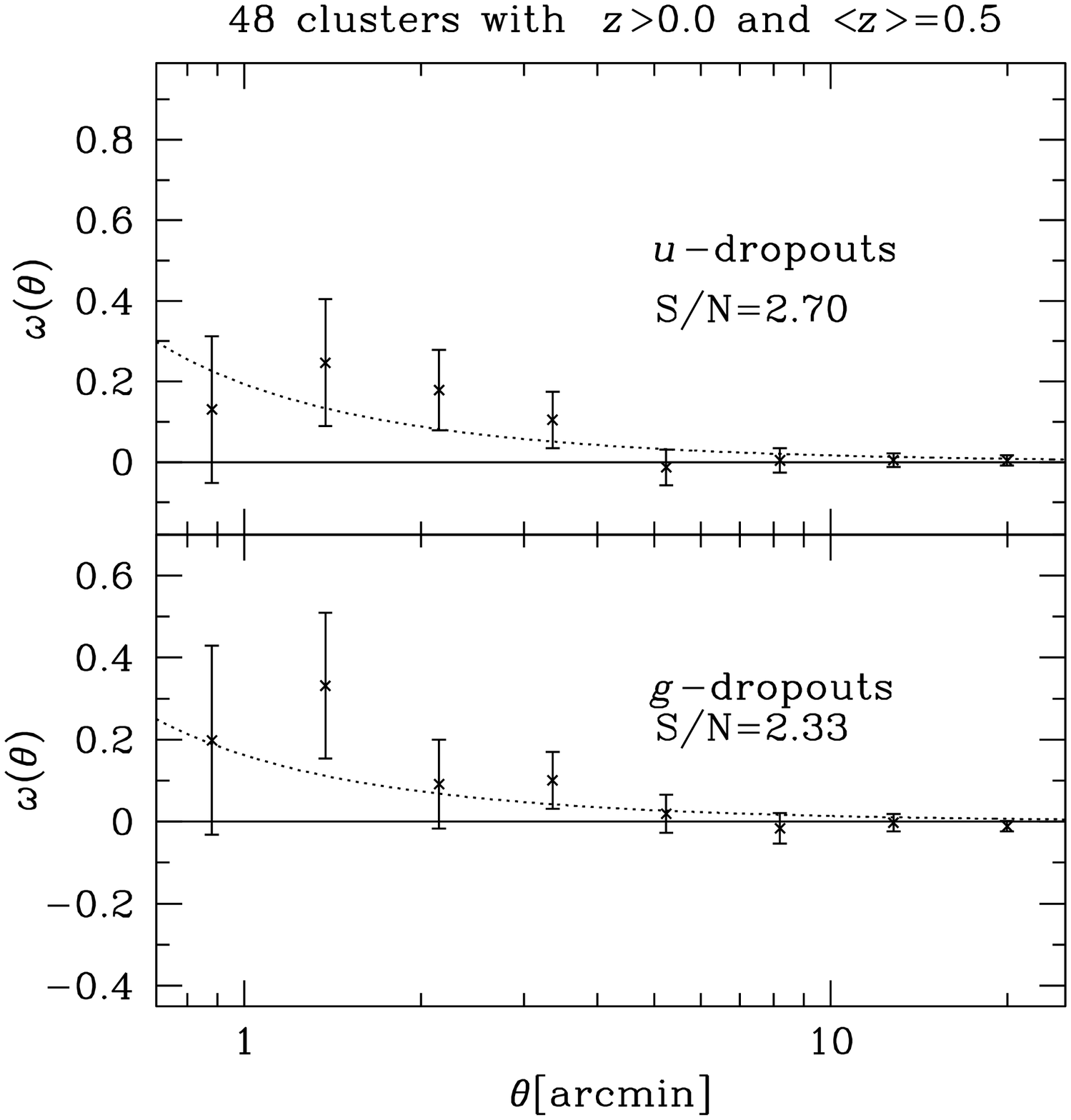}
\includegraphics[width=0.32\textwidth]{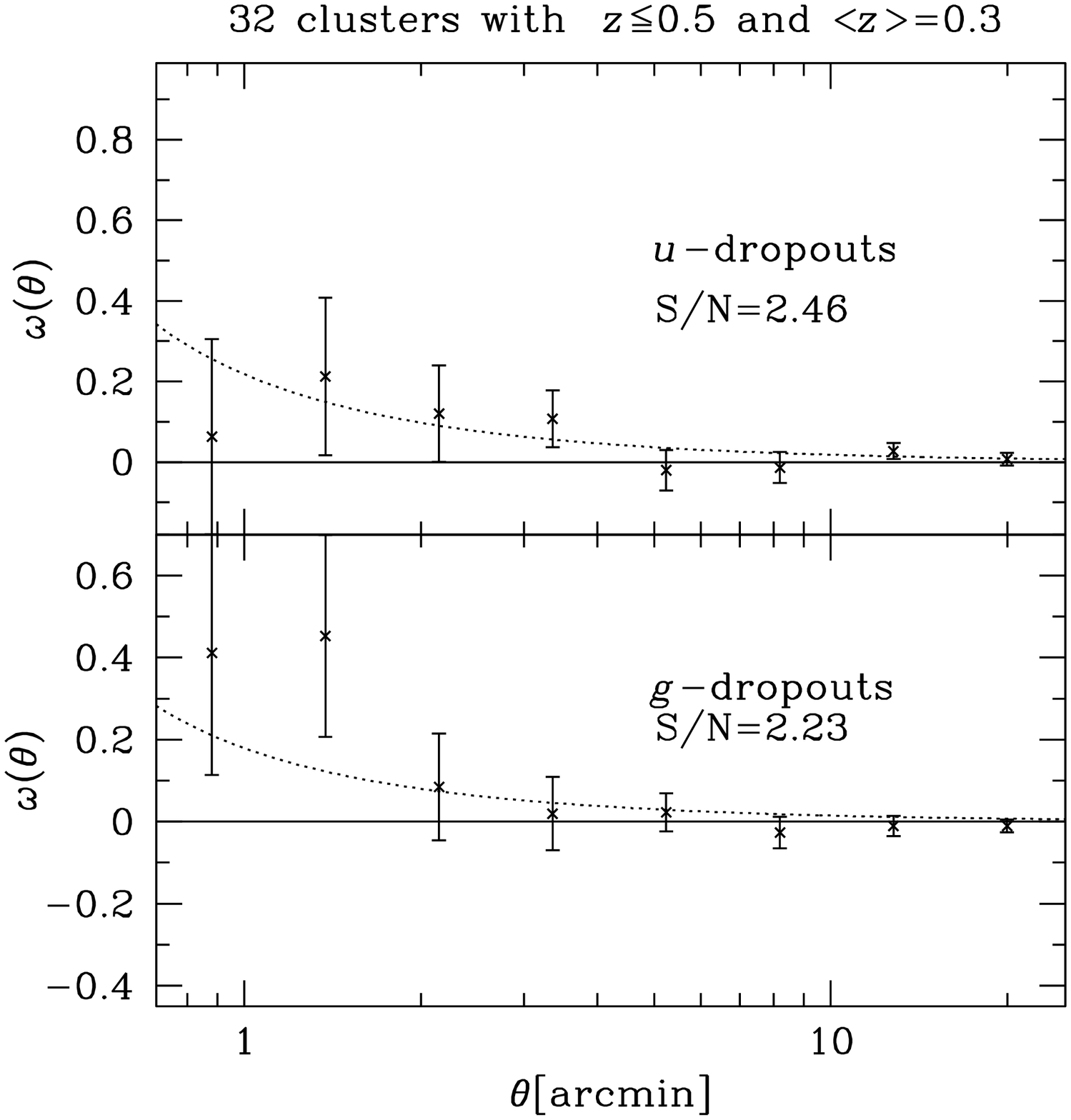}
\includegraphics[width=0.32\textwidth]{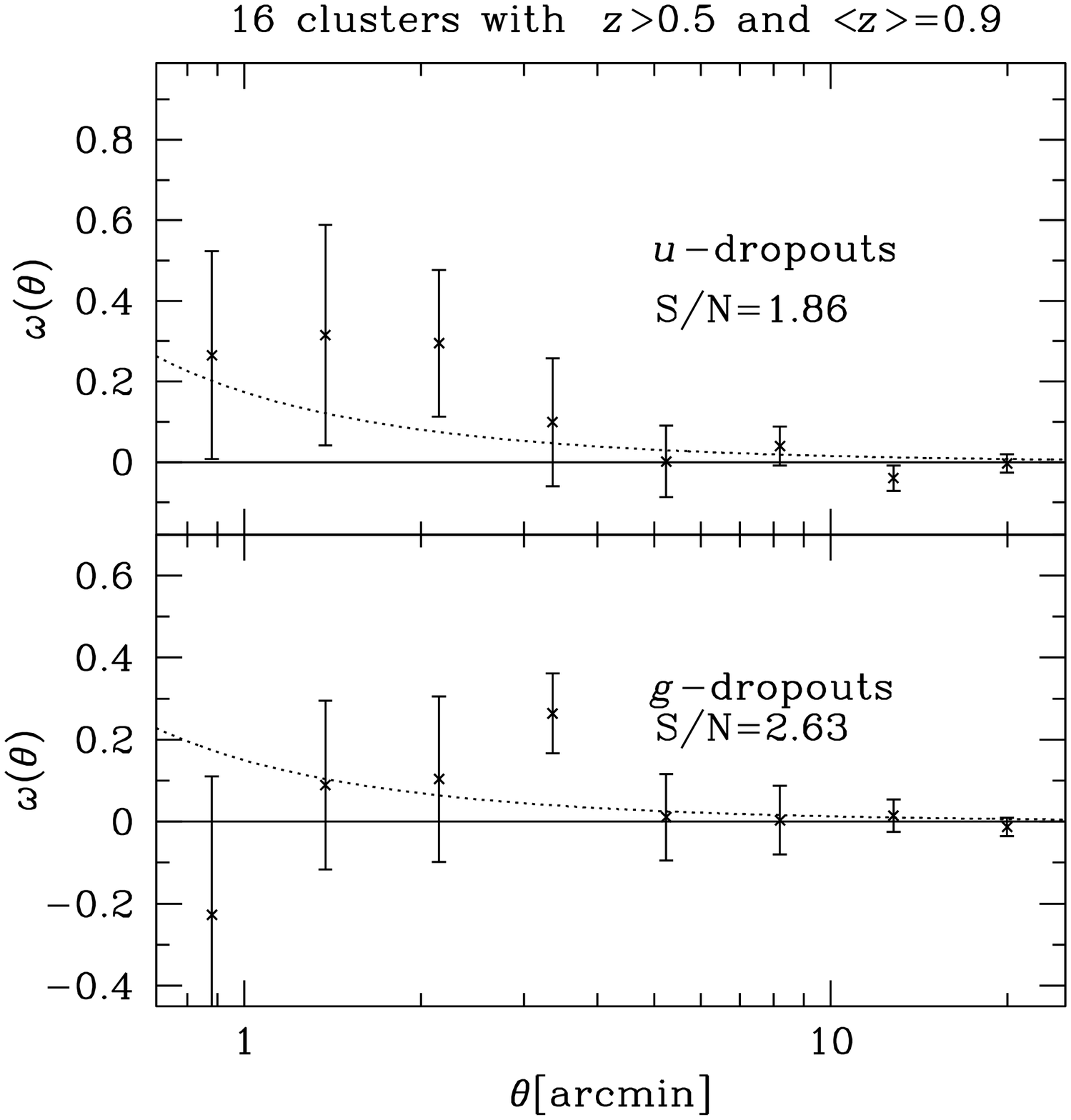}
\caption{Magnification-induced cross-correlation between rich SpARCS
  cluster and $u$-dropouts (\emph{top}) and $g$-dropouts
  (\emph{bottom}) from CFHTLenS. The dotted lines correspond to the
  best-fit multi-SIS profiles and the S/N values are for the
  individual measurements (see Table~\ref{tab:results} for the S/N
  when $u$- and $g$-dropouts are combined). \emph{Left:} All 48
  clusters. \emph{Middle:} Subsample of 32 clusters with
  $z\le0.5$. \emph{Right:} Subsample 16 clusters with $z>0.5$.}
\label{fig:signal}
\end{figure*}

Fig.~\ref{fig:signal} shows the cross-correlation signal between rich
SpARCS clusters in two different redshift bins and the $u$-dropouts
with $23<r<24.5$ ($\left<\alpha_{\rm eff}-1\right>=1.54$; surface
density $0.26 \,{\rm arcmin}^{-2}$) and $g$-dropouts with $23.5<i<25$
($\left<\alpha_{\rm eff}-1\right>=1.24$; surface density $0.24 \,{\rm
  arcmin}^{-2}$).\footnote{When comparing surface densities between
  shear and magnification to estimate the S/N, the magnification
  densities should be scaled by a factor $\left(2 |\alpha_{\rm eff}-1|
  \sigma_{\epsilon}\right)^2$ \citep{2000A&A...353...41S}, with
  $\sigma_\epsilon$ being the intrinsic ellipticity dispersion.}  The
mean masses of the clusters are estimated from their richness. Errors
on the data points represent the standard deviation between the
signals for the different clusters and thus reflect Poissonian noise
as well as noise due to clustering of the background population and
noise introduced by structures along the lines-of-sight to the
clusters \citep{2011MNRAS.tmp...72H}. The off-diagonal elements of the
covariance matrix are estimated by measuring the cross-correlation
signal between 1000 fake clusters placed at random positions and the
background LBGs.

We fit Multi-SIS (see Eq.~\ref{eq:multi_SIS}) profiles to the combined
$u$- and $g$-dropout signals for the two cluster samples in the range
$0\farcm8<\theta<20'$. Table~\ref{tab:results} summarizes the
results. The WL signal is - within the large errors - consistent with
the signal expected from the \cite{2007ApJ...663..150M}
  mass-richness relation, as indicated by the values for $a$ being
  consistent with unity. Despite similar S/N the relative error of
the mass is larger for the high-$z$ sample than for the low-$z$
sample. That is due to the fact that the high-$z$ $g$-dropout signal
is not very well-fit by the multi-SIS profile (but still consistent
within noise). Note that we just used the \emph{scaling} with richness
to construct the multi-SIS function and not the absolute
\emph{normalization} of the mass-richness relation. Separate multi-SIS
fits to the signals of the $u$- and $g$-dropouts of a given cluster
sample yield consistent results as well. In the following we perform
two tests for systematic errors that could affect our result.

\begin{table}
\caption{Results of the multi-SIS fits to the magnification signals.}
\begin{tabular}{cccccccc}
\hline
\hline
sample & \# cl. & $\left< z\right>$ & $M_{\rm richness}$ & S/N & $a$ & $M_{\rm magn.}$\\
&&& $\times10^{14}M_\odot$&&& $\times10^{14}M_\odot$\\
\hline
all       & 48 & 0.5 & 1.9 & 3.6 & $1.1^{+0.6}_{-0.6}$ & $2.1^{+1.2}_{-1.1}$\\
$z\le0.5$ & 32 & 0.3 & 1.1 & 3.3 & $1.4^{+0.8}_{-0.7}$ & $1.5^{+0.9}_{-0.8}$\\
$z>0.5$   & 16 & 0.9 & 3.4 & 3.2 & $0.9^{+0.9}_{-0.7}$ & $3.1^{+3.0}_{-2.5}$\\
\end{tabular}
\label{tab:results}
\end{table}

The SpARCS XMM-LSS field also overlaps with the CFHTLS-Deep field
D1. In this one deg$^2$ field much deeper imaging is available
which allows the selection of fainter LBGs \citep[the ones also used
  in][]{2009A&A...498..725H,2009A&A...507..683H} which have negative
$\left<\alpha_{\rm eff}-1\right>$. We find eight of the 48 clusters in
that field. Although the statistical power is limited due to
the small number of clusters and the lower absolute value of
$\left<\alpha_{\rm eff}-1\right>$ (somewhat compensated by the higher
LBG density), these faint LBGs allow us to perform an important
test. While the positive cross-correlations of the brighter LBGs could
in principle also be caused by improper redshift separation and hence
physical cross-correlations, anti-correlations can only be caused by
WL. Here we use faint $u$- and $g$-dropouts with $r>25.8$ and
$i>26.3$, respectively, which correspond to parts of the LBG
luminosity function where $\left<\alpha_{\rm eff}-1\right>=-0.09$ and
$\left<\alpha_{\rm eff}-1\right>=-0.18$. Their surface densities are
$2.7 \,{\rm arcmin}^{-2}$ and $0.83\,{\rm arcmin}^{-2}$.

As expected for a background sample with $\left<\alpha_{\rm
  eff}-1\right><0$, we find under-densities of LBGs around the
clusters. The detection is less significant due to the limited number
of clusters and the lower absolute values of $\left<\alpha_{\rm
  eff}-1\right>$ which cannot be compensated by the higher surface
density of dropouts in the CFHTLS-Deep and their lower intrinsic
clustering signal.\footnote{The main noise term in the magnification
  measurement is not the Poissonian shot-noise, which originates from
  the finite sampling of the $\kappa$-field with a limited number of
  background objects, but the noise introduced by the intrinsic
  clustering of the background galaxies \citep[see][for a full
    treatment of the different error terms]{2010MNRAS.401.2093V}. As
  we showed in \cite{2007A&A...462..865H,2009A&A...498..725H} the
  clustering of the faint LBGs is much weaker than the clustering of
  the bright LBGs.}

Another useful test is to cross-correlate the clusters to a population
of objects that should not show any correlation in their positions
because it is closer to the observer than the clusters. For that test
we select stars and measure their cross-correlation function to the
clusters. We do not detect any significant cross-correlation signal
for stellar samples of different limiting magnitudes. It should be
noted that the stellar auto-correlation function is different from
zero (with a very low amplitude, but high significance) so that a
non-detection of a cross-correlation signal between stars and clusters
is a meaningful test.

From these two tests we conclude that our measurement is robust and
free of strong systematic biases. However, the current sample is two
small to study more subtle systematic effects in detail.

\section{Discussion and Outlook}
\label{sec:discussion}
We measure the normalization of the high-redshift mass-richness
relation of galaxy clusters selected from SpARCS employing a novel
method based on the WL magnification effect of these clusters on
$z\ga3$ LBGs. Even with a small sample of rich, high-$z$ clusters and
optical data of modest depths we detect a magnification signal at
$>3$-$\sigma$ significance which yields a measurement of this
normalization (equivalent to a measurement of the average cluster
mass) with a relative error of $\sim 50\%$.\footnote{Note that the
  relative error of $a$ scales as $\left({\rm S\over N}\right)^{-2/3}$
  for data that agree within errors with the model.} The measured
normalization is in good agreement with the low-$z$ mass-richness
relation from \cite{2007ApJ...663..150M} which is based on dynamical
mass estimates.

Using much deeper data on a one deg$^2$ sub-field we detect an
anti-correlation between a few high-$z$ SpARCS clusters and faint LBGs
supporting a WL origin of the signal and suggesting that the
measurement on the shallower data is free of physical
cross-correlations that could mimic a WL signal.

The overlap area between SpARCS and CFHTLS is too small to subdivide
the clusters into different richness bins and estimate a full
mass-richness relation at high redshift. However, it is well within
reach of current facilities to expand this area by an order of
magnitude through either observing IR-, X-ray-, or SZ-surveys with
ground-based optical imaging or through targeting optical surveys such
as the CFHTLS with IR- or X-ray imaging from space or with SZ
telescopes.

We would like to stress that this method represents a way to obtain
mass estimates of large samples of clusters at very high redshifts
that is relatively cheap in terms of telescope time. It is the only
method that can realistically be applied to measure the masses of very
large numbers of $z>1$ clusters to be found by future high-$z$ cluster
surveys. Concentrating on the bright part of the LBG luminosity
function and probably also on the lowest redshift and hence brightest
LBGs reachable from the ground ($u$-dropouts) seems the most promising
route. This certainly requires high-quality $u$-band imaging which is
available at a few facilities only.

\acknowledgments

Based on observations obtained with MegaPrime/MegaCam, a joint project
of CFHT and CEA/DAPNIA, at the CFHT which is operated by the NRC of
Canada, the CNRS of France, and the University of Hawaii and on
observations made with the Spitzer Space Telescope, which is operated
by the Jet Propulsion Laboratory, California Institute of Technology
under a contract with NASA. This work is based in part on data
products produced at TERAPIX and CADC.

We would like to thank the CFHTLenS team for their work on the
CFHTLenS data products which are used in this study. H. Hildebrandt is
supported by the Marie Curie IOF 252760 and by a CITA National
Fellowship. TE is supported by the BMBF through project "GAVO III" and
by the DFG through project ER 327/3-1 and the TR 33. H. Hoekstra
acknowledges support from the NWO and a Marie Curie IRG. LVW is
supported by NSERC and CIfAR. GW acknowledges support from NSF grant
AST-0909198.

Facilities: \facility{Spitzer}, \facility{CFHT}

\end{document}